%% file: main.tex
\documentclass[11pt,a4paper]{article}
\usepackage{jinstpub}	
\pdfoutput=1 	

\usepackage{graphicx} 	
\usepackage{amssymb} 	
\usepackage{mathtools} 	
\usepackage{comment} 	
\usepackage{color} 		
\usepackage{multirow}	
\usepackage{gensymb}
\usepackage{subcaption}
\usepackage{epstopdf}
\usepackage[]{siunitx}  
\DeclareSIUnit\sample{S}
\DeclareSIUnit\bits{bits}

\usepackage{lineno} 	
\newcommand*\patchAmsMathEnvironmentForLineno[1]{%
  \expandafter\let\csname old#1\expandafter\endcsname\csname #1\endcsname
  \expandafter\let\csname oldend#1\expandafter\endcsname\csname end#1\endcsname
  \renewenvironment{#1}%
     {\linenomath\csname old#1\endcsname}%
     {\csname oldend#1\endcsname\endlinenomath}}%
\newcommand*\patchBothAmsMathEnvironmentsForLineno[1]{%
  \patchAmsMathEnvironmentForLineno{#1}%
  \patchAmsMathEnvironmentForLineno{#1*}}%
\AtBeginDocument{%
\patchBothAmsMathEnvironmentsForLineno{equation}%
\patchBothAmsMathEnvironmentsForLineno{align}%
\patchBothAmsMathEnvironmentsForLineno{flalign}%
\patchBothAmsMathEnvironmentsForLineno{alignat}%
\patchBothAmsMathEnvironmentsForLineno{gather}%
\patchBothAmsMathEnvironmentsForLineno{multline}%
}

\usepackage{xspace} 	
\usepackage[all]{hypcap} 

\newcommand{\znbb} {$0\nu\!\beta\!\beta$\xspace}

\newcommand{\otsx} {$^{136}\mathrm{Xe}$\xspace}

\newcommand{\sep}{, } 
\newcommand{\Efield} {936~V/\si{cm}\xspace}

\newcommand{\DriftL}{33.2~\si{mm}\xspace}
\newcommand{\DriftTL}{16.6~\si{\us}\xspace}
\newcommand{\DriftS}{18.2~\si{mm}\xspace}
\newcommand{\DriftV}{2~\si{\mm/\us}\xspace}
\newcommand{\DataRes}{5.5\% \xspace}
\newcommand{\BiPeakL}{570~\si{keV}\xspace}
\newcommand{\BiPeakH}{1064~\si{keV}\xspace}
\newcommand{\XeMass}{9~\si{kg}\xspace}
\newcommand{\CathBias}{-3110~\si{V}\xspace}
\makeatletter
\renewcommand\tableofcontents{%
    \@starttoc{toc}%
}
\makeatother

\newif\ifshowchanges
\showchangesfalse

\newcommand{\vo}[1]{}

\ifshowchanges
\usepackage{soul}
\soulregister\cite7
\soulregister\ref7
	\renewcommand{\vo}[1]{{\textcolor{red}{\st{#1}}}}
    
\fi

\begin{document}

\title{Characterization of an Ionization Readout Tile for nEXO}
\input{authors_full.tex}
\arxivnumber{1710.05109}
\date{\today}

\abstract{%
A new design for the anode of a time projection chamber, consisting of a charge-detecting "tile", is investigated for use in large scale liquid xenon detectors. The tile is produced by depositing 60 orthogonal metal charge-collecting strips, 3~mm wide, on a 10~\si{\cm} $\times$ 10~\si{\cm} fused-silica wafer. These charge tiles may be employed by large detectors, such as the proposed tonne-scale nEXO experiment to search for neutrinoless double-beta decay.  Modular by design, an array of tiles can cover a sizable area.  
The width of each strip is small compared to the size of the tile, so a Frisch grid is not required. A grid-less, tiled anode design is beneficial for an experiment such as nEXO, where a wire tensioning support structure and Frisch grid might contribute radioactive backgrounds and would have to be designed to accommodate cycling to cryogenic temperatures. The segmented anode also reduces some degeneracies in signal reconstruction that arise in large-area crossed-wire time projection chambers.  A prototype tile was tested in a cell containing liquid xenon. Very good agreement is achieved between the measured ionization spectrum of a $^{207}$Bi source and simulations that include the microphysics of recombination in xenon and a detailed modeling of the electrostatic field of the detector. An energy resolution $\sigma/E$=\DataRes is observed at \BiPeakL, comparable to the best intrinsic ionization-only resolution reported in literature for liquid xenon at \Efield.
}


\maketitle


\section{Introduction}
\label{sec:sec-intro}

The search for neutrinoless double beta (\znbb) decay is an active field of research.  The observation of this lepton-number-violating process would indicate that neutrinos are Majorana particles and could constrain the absolute neutrino mass scale~\cite{Dell'Oro:2016dbc}.  The planned next-generation detector (nEXO) proposes to search for \znbb decay of $^{136}$Xe  in a 5~tonne liquid xenon (LXe) time projection chamber (TPC)~\cite{nEXOSen2017}. The nEXO experiment plans to build on the success of the currently operating EXO-200 experiment~\cite{exo200upgrade}. 

In EXO-200 the ionization signal is measured by two planes of crossed wires where one plane is used as a shielding grid and the other as the charge collection grid~\cite{Auger:2012gs}.  The EXO-200 TPC is approximately 36~cm in diameter while the diameter of nEXO will be over 1~\si{m}. At this larger diameter, a rather substantial tensioning frame that can be temperature cycled to $\sim$165~\si{K} would be required, which would pose a challenge to the required radioactivity budget of nEXO.  In addition, the larger its diameter, the more vulnerable a crossed-wire design is to ambiguity in reconstructing the position of multiple energy deposits in the detector.  Wires are also susceptible to microphonic pickup from environmental noise.  For these reasons it was suggested \cite{sinclair2012} to explore using anode pads as an alternative readout.

To avoid issues with long crossed wires, the nEXO collaboration is investigating a segmented anode composed of an array of tiles. Each tile consists of a dielectric substrate covered with an array of conductive strips for collecting charge. The channel segmentation offered by a tiled design strongly mitigates the likelihood of ambiguity in the reconstructed position of the charge deposition. As described in Section~\ref{sec:app}, the charge tiles can be made using only materials that are either known to be obtainable with extremely low radioactive contamination (such as fused silica), or employed in minimal amounts (i.e.\ the thin conductive strips). Finally, no mechanically-robust tensioning system is required.

In this paper, a prototype tile is tested in an LXe TPC, and results are compared to simulation. The observed performance is compared to that of more traditional designs.  

\section{Experimental Apparatus}
\label{sec:app}

\subsection{Prototype Anode Tile}
\label{sec:apptile}
A 10~\si{\cm} $\times$ 10~\si{\cm} prototype ionization readout tile which is 300~\si{\um} thick was fabricated by the Institute of Microelectronics of the Chinese Academy of Sciences. The tile substrate is a fused-silica wafer with 60 electrically isolated strips (30 "X" strips and 30 orthogonal "Y" strips).  Strips are made by depositing layers of Au and Ti onto the silica wafer surface.

\begin{figure}
    \centering
   \includegraphics[trim={1cm 0 0.5cm 0cm},clip, width=\linewidth]{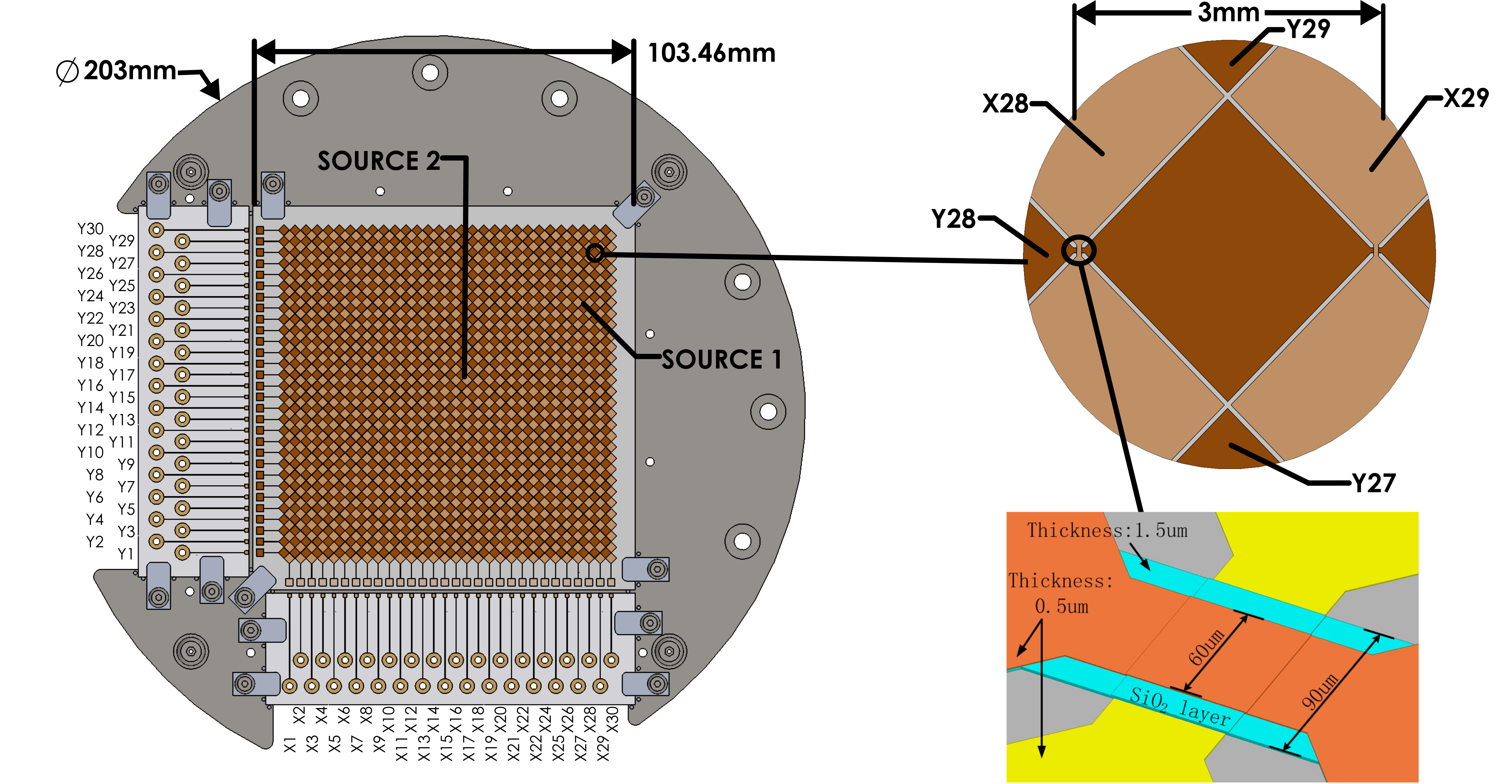}
    \caption{Sketch of the prototype charge tile (left), with 60 orthogonal strips, 30 in each direction mounted on a stainless steel support used for testing. The light-colored X strips are vertical and the darker Y strips are horizontal in this view.  Each strip is made of 30 3~\si{\mm} $\times$ 3~\si{\mm} square pads connected at two opposite corners. Also shown (top right) is a detail view of strips on the anode tile.  The X and Y strips cross each other at the pad junctions (bottom-right).  The metal X and Y layers are separated by a thin layer of SiO$_{2}$. This arrangement maximizes overall pad coverage while limiting the capacitance.  The projected locations of the two $^{207}$Bi sources located on the opposite cathode are indicated.  Also shown are the ceramic interface boards mounted on two sides of the tile for testing purposes. Pads on the ceramic interface allow wire bonding to the tile. The ring terminals are connected to feedthrough leading to pre-amplifiers outside of the LXe.} 
    \label{Tilefig1}
\end{figure}

Each strip is approximately 10~cm long consisting of 30 square pads, which are 3~\si{\mm} across the diagonal and daisy-chained at their corners.  This geometry maximizes the metallic cover of the substrate, reducing the risk of charge accumulation, and minimizes the capacitance at the crossing between strips.  Layers of 1.5\si{\um} thick SiO$_2$ are used at the crossing points of X and Y strips to provide electrical isolation. The capacitance at each crossing is 80~\si{\fF}, assuming that the conducting structures are 0.5~\si{\um} thick gold.  This results in a capacitance of 0.57~\si{\pF} between pairs of crossed strips and a capacitance of 0.86~\si{\pF} between adjacent parallel strips.  In addition a resistance of $\sim$5~\si{\ohm} at LXe temperature is expected along each single strip on the tile.  A diagram of the tile mounted on a stainless steel support used for testing is shown in Figure~\ref{Tilefig1} with details of the strip geometry and orthogonal strip crossing in the insets.

\begin{figure}
    \centering
   \includegraphics[trim={0cm 0 0cm 0cm},clip, width=0.9\linewidth]{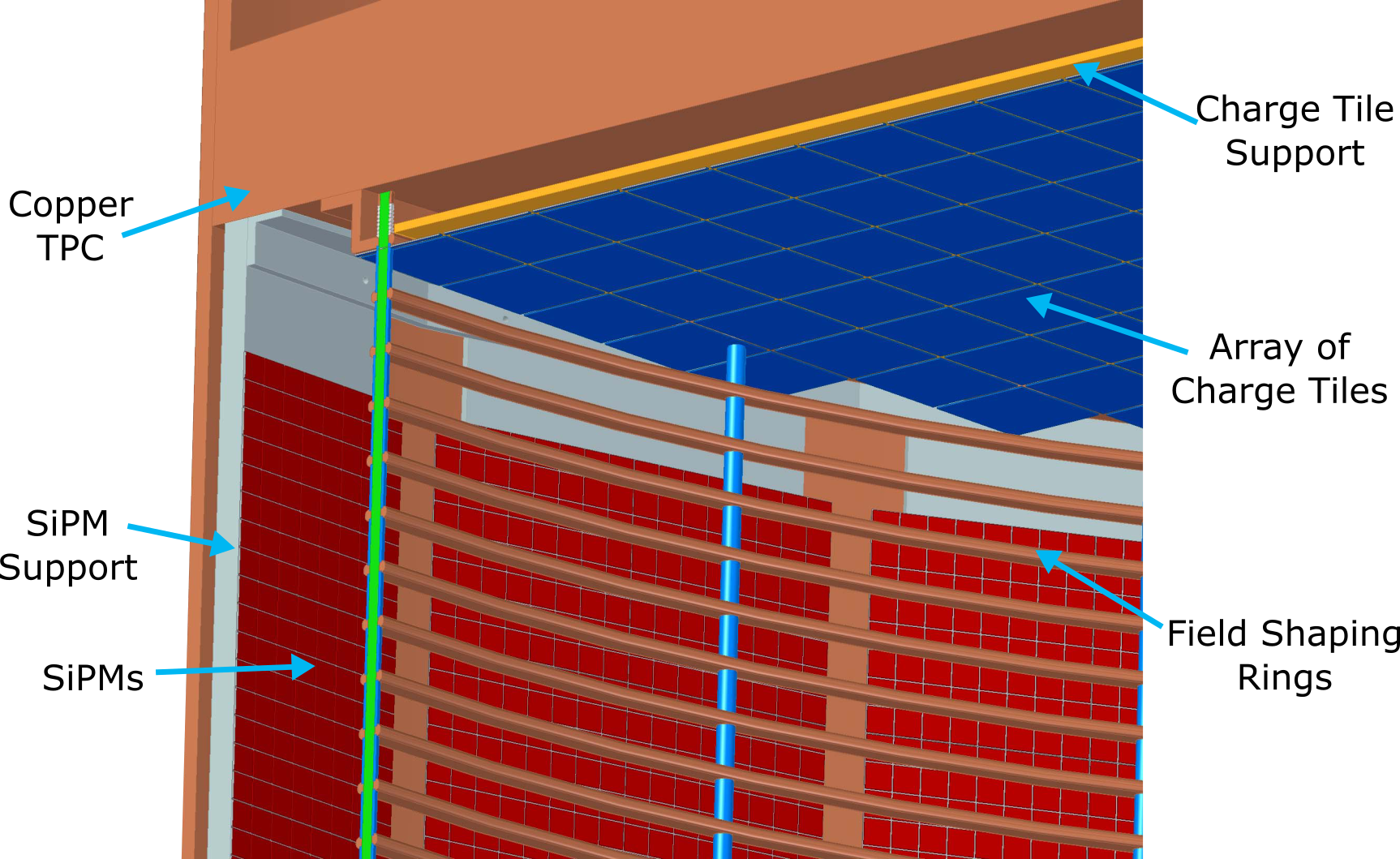}
    \caption{Diagram of the integration scheme of charge tiles into the nEXO LXe TPC.  The anode is composed of a mosaic of many adjacent charge tiles.  Also shown on the outer edge of the copper TPC are UV-sensitive silicon photomultipliers (SiPMs), the proposed method for detecting the LXe scintillation light.} 
    \label{fig:nEXOTile}
\end{figure}

The design of the tile tested here is representative of what is currently proposed for nEXO, although parameter optimization is still under way. The integration concept for the nEXO charge collection plane composed of many such charge tiles covering the anode surface is shown in Figure~\ref{fig:nEXOTile}. This represents a departure from the crossed wire plane design adopted by EXO-200 and avoids the need to provide a substantial tensioning frame that is both cryogenic compatible and meets the required radioactivity budget.  This approach does not use a Frisch grid but offers the advantage of additional channel segmentation to reduce possible ambiguity in reconstructing the position of individual charge clusters in events with multiple charge depositions.

\subsection{Test Cell}
\label{sec:appcell}

\begin{figure}
    \centering
    \includegraphics[trim={0cm 0cm 0cm 0cm},clip,width=\linewidth]{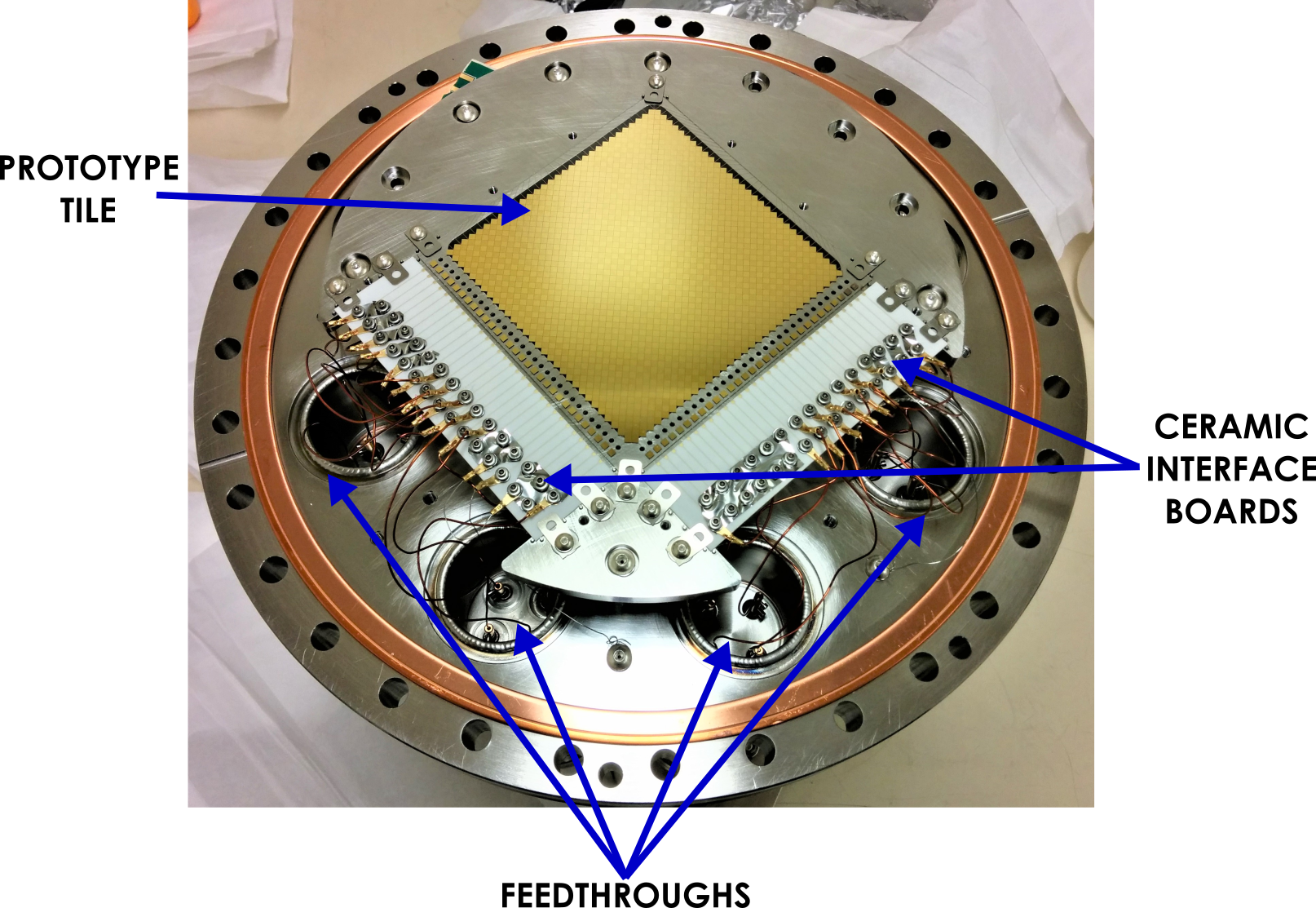}
    \caption{Photograph of the anode tile mounted to the top flange of the TPC.}
    \label{TPCpic}
\end{figure}

\begin{figure}
    \centering
    \includegraphics[trim={2.2cm 0 1.2cm 0cm},clip,width=\linewidth]{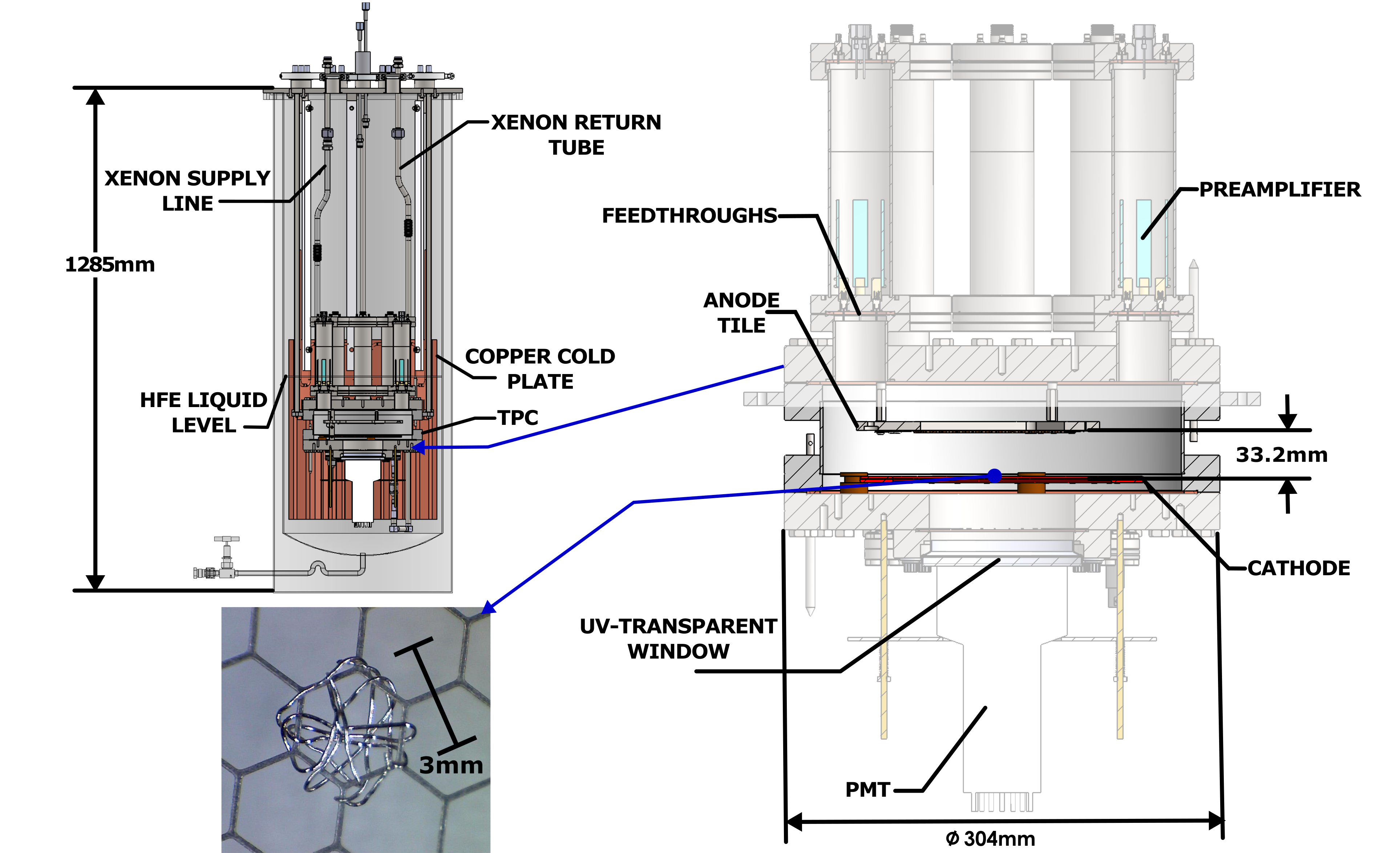}
    \caption{Sketch of the TPC mounted inside the HFE-7000 dewar (top left) showing the key cryogenic components.  Also shown is a cut out view of the TPC (right) and a photograph of one of the Bi$^{207}$ sources woven into the cathode mesh (bottom left).}
    \label{fig:dewar}
\end{figure}

In this test setup, the anode mounting plate is inserted into a liquid xenon time projection chamber (TPC) as shown in Figure~\ref{TPCpic} and Figure~\ref{fig:dewar}.  This TPC is used to characterize the performance of prototype anode tiles in LXe. The TPC is built from a 12~\si{in} ConFlat spool-piece and two flanges. The body of the TPC is 304 stainless steel with a 30~\si{cm} outer diameter and a 13~\si{\cm} height. Both TPC anode and cathode mounting plates are approximately 20~\si{\cm} in diameter.  The prototype tile is mounted at the center of the anode plate as shown in Figure~\ref{Tilefig1}.  The drift length can be varied from \DriftS to \DriftL by changing spacers behind the anode, but was set to the longer length of \DriftL for the data presented here.

On the tile, each strip ends with a square pad where the signal can be read out. For this test each channel is wire-bonded to an adjacent ceramic interface board which is also mounted on the anode plate (Figure~\ref{TPCpic}). Ring terminals crimped to Kapton-insulated copper signal wires are bolted to through holes on the ceramic interface boards and bring the signals to electrical feedthroughs on top of the xenon cell (Figure~\ref{fig:dewar}). This solution allows for easy reconfiguration of the strip-readout channel assignments and avoids the use of solder, which could deteriorate the chemical purity of the LXe.

The center of the TPC cathode, 150~\si{\mm}  in diameter, is a photo-etched 302 stainless steel hexagonal mesh, 127~\si{\um} thick with 95\% optical transparency and 3~\si{\mm} hexagon size (see Figure~\ref{fig:dewar}). Below the cathode is a 6~\si{in} diameter optical window made with UV grade synthetic silica.  Clamped against the face of the optical window and located outside of the xenon volume submerged in the HFE is a cryogenic photomultiplier (PMT), EMI 9921QB, optimized for VUV sensitivity. The PMT collects the xenon scintillation light (peaked at 178~\si{nm}) and provides a trigger signal for each event. The cathode is typically biased at \CathBias resulting in an average drift field of \Efield. 

The TPC is filled with $\sim$\XeMass of LXe, which is sufficient to fully submerge the anode tile.  Operation of the detector is typically performed at $\sim$900~torr and $\sim$168~K. Mimicking the cryogenic scheme used in EXO-200 and proposed for nEXO, the TPC is submerged in HFE-7000 cooling fluid \cite{hfe} to maintain good temperature uniformity.  The HFE-7000 is contained in a cryostat and cooled via a large copper heat exchanger ("cold plate") immersed in the HFE-7000 and cooled with liquid nitrogen flowing through internal tubing.  The cold plate is shaped as a semi-cylindrical surface that wraps around the TPC chamber, purposely azimuthally asymmetric to force convection of the HFE-7000 fluid.  The temperature of the cold plate is measured by three thermocouples placed at different locations on its body.  One of them regulates the flow of the liquid nitrogen via a PID feedback loop managed by a LabVIEW  application \cite{labview}.  The temperature of the cell is recorded using additional thermocouples attached at three locations on the outside of the TPC.  The pressure is also monitored at two locations in the Xe system using two Model 121A MKS Baratrons \cite{mks_bara}. The HFE-7000 dewar and TPC are shown in Figure~\ref{fig:dewar}.

Before liquefaction the TPC is initially pumped down to $\lesssim 10^{-6}$~\si{torr}, which results in a small enough oxygen contamination to achieve the required LXe purity for drifting electrons over the TPC length. The empty cell and the HFE-7000 are cooled down and held at the base operating temperature of 168~\si{K}. When filling the detector, the xenon is passed through a SAES MonoTorr model PF3C3R1 getter \cite{saes_getter} to remove electronegative impurities and transferred into the cold cell, where it enters through a line mounted on the side and liquefies.  Continuous recirculation through the purifier is not currently needed since there is no noticeable degradation in the electronegative purity during the typical data taking run ($\sim$2~days).  A limit of $\sim$150~\si{\us} on the electron lifetime is estimated by observing the variations in the location of the \BiPeakL peak as a function of drift time in data and simulation.   
This lifetime is sufficient to drift and collect charge in the current setup, which has a maximum drift time of \DriftTL.  In nEXO the expected electron lifetime is  $>$10~\si{ms} but the ultimate drift length in nEXO is expected to be $\sim$625~\si{\us} \cite{nEXOSen2017} which is significantly larger than that achieved here.  The approximate ratio of lifetime to drift length is still comparable making this representative of the expected purity effects in nEXO. A MKS 1479A Mass-Flow Meter \cite{massflow} mounted along the room-temperature portion of the fill line is used to determine the volume of xenon flowing into the cell.     

A $^{207}$Bi source is plated onto a platinum-rhodium wire, 127~\si{\um} in diameter and woven onto the cathode mesh in two locations as shown in Figure~\ref{fig:dewar}.  Projected onto the anode tile the first source is approximately located at the crossing of channels X28 and Y24 and the second at the crossing of channels X17 and Y17, as shown in Figure~\ref{Tilefig1}. Each source emits two primary gamma rays of \BiPeakL and \BiPeakH.  The event rate (before any veto) measured as the PMT trigger rate is 960~Hz. Each source contributes approximately half of the activity.

\section{Data Taking}
\label{sec:data}

\begin{figure}
\centering
\includegraphics[width=0.98\linewidth]{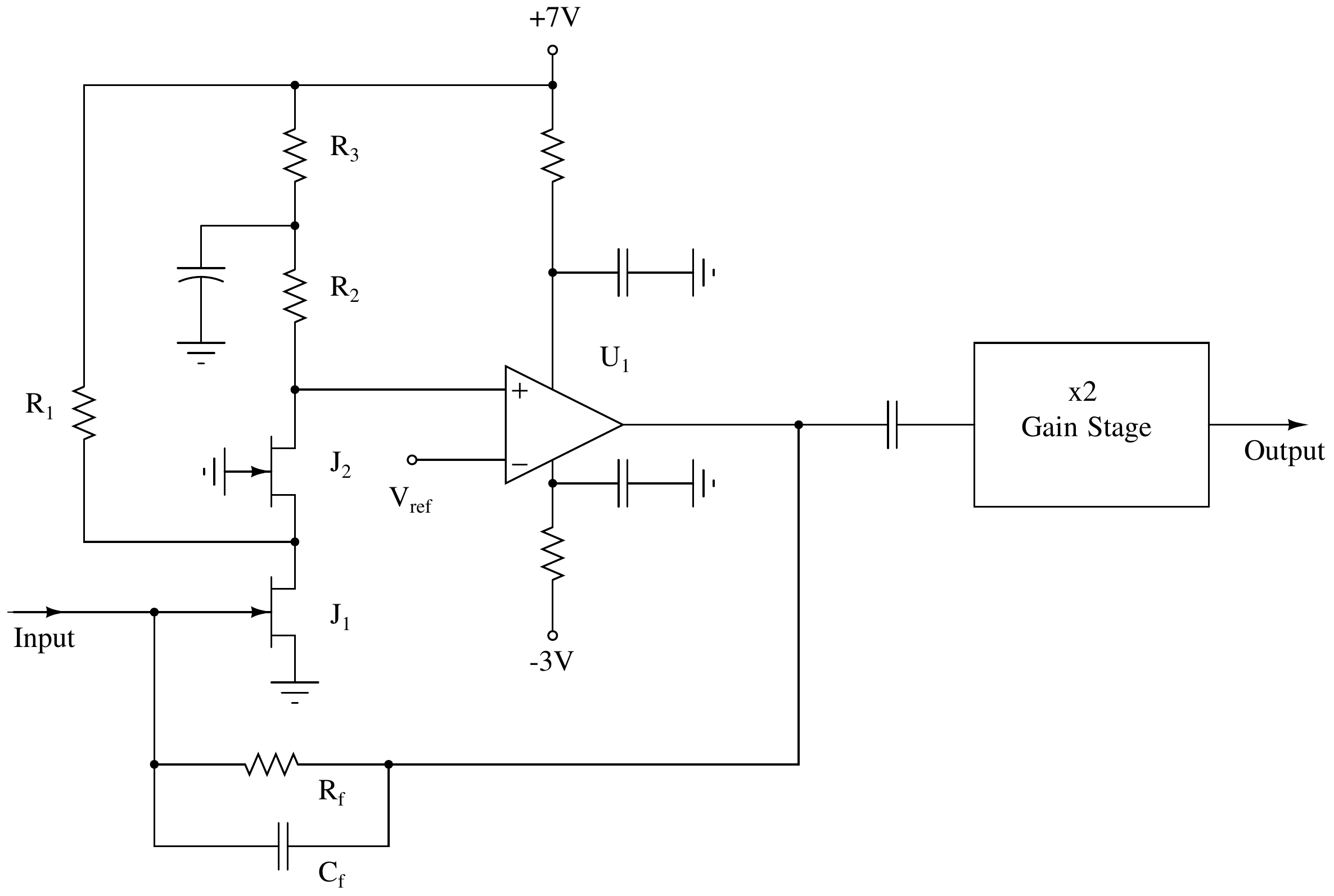}
\caption{Schematic diagram of the charge-sensitive preamplifiers used in this study. The input JFET, J1, is a commercial off-the-shelf BF862, at about 5 mA drain current. The current is set mostly by the series R2 and R3, and is boosted as needed by R1. The transistor J2 acts as a cascode to improve dynamic performance and increase the open-loop gain at medium to high frequencies. The presence of the operational amplifier U1 guarantees a very large value for the open-loop gain of the circuit to minimize potential crosstalk among adjacent strips. The second stage of gain is optional, but in this case was used to match the DAQ dynamic range.}
\label{fig:preamp}
\end{figure}

Each charge-readout channel from the  tile is instrumented with a custom charge-sensitive preamplifier, based on the architecture shown in Figure~\ref{fig:preamp}.  This architecture was chosen because of its low noise characteristics (200 electrons RMS at a peaking time greater than 3~\si{\us} for 24 pF strip capacitance and an operation temperature of $\sim$168~\si{K}), compact footprint, and the ability to drive cables over long distances. Power consumption was not a factor, nevertheless, each preamp meets the electrical specifications at less than 45~\si{mW} power consumption. An additional stage of gain is placed after the preamplifier in order to match the dynamic range of the signal processing electronics. The full functionality of the preamplifier is described in~\cite{Fabris1999545}. 

The preamplifiers are installed in eight leak-tight cans filled with nitrogen gas and mounted just above the TPC feedthroughs to minimize the distance from the strips to the front-end electronics, which in our case is approximately 10~\si{\cm}.  The cans are partially submerged in HFE to keep them near the designed operating temperature of $\sim$168~\si{K}.  No temperature sensors are installed to measure the temperature inside the cans so the exact operating temperature is not precisely known and the electronic noise is expected to grow proportionally to the temperature.  The output of each preamplifier is further amplified 10-fold by a Phillips Scientific 776 amplifier installed $\sim$2~\si{m} away and outside the cryostat.  The output signal from the second stage of amplification is then sent to a digitizer were the waveforms are readout and stored on the DAQ computer. 

Digitized waveform data is collected at 25~MS/s with two 16-channel 14-bit SIS3316 digitizers~\cite{StruckDigi}. Thirty two channels are recorded: thirty charge channels, the PMT signal, and a pulser signal used for noise measurements.  The signal from the PMT is split into two: one signal enters a discriminator to generate an event trigger, the other signal is shaped with an Ortec~672 spectroscopy amplifier with 500~\si{\ns} shaping time, then digitized.  Since there are sixty charge channels on the tile and only thirty digitizer channels available, some strips are ganged together on the ceramic interface board inside the TPC. Data from all 32 readout channels is collected during each event. A prescaler vetoes 4.3~ms after every PMT trigger, to keep the data to a rate of about 1~GB per minute, the maximum that is manageable for the data acquisition computer. 

\begin{figure}
    \centering
    \includegraphics[trim={0.8cm 0 2.3cm 0.5cm},clip,width=\linewidth]{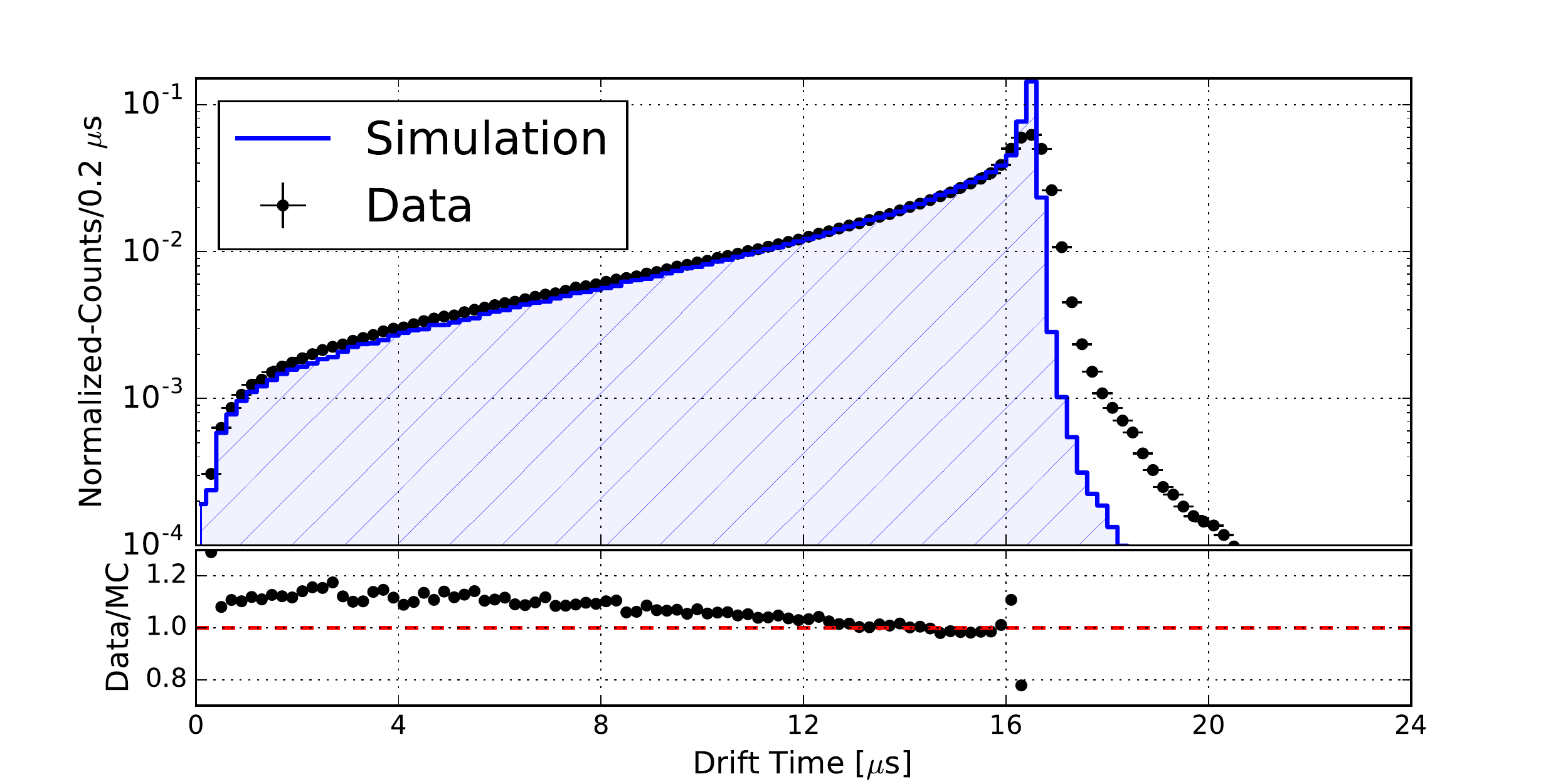}
    \caption{Spectrum of drift times in both data (points) and simulation (shaded area) with the ratio of data to simulation shown in the lower panel. The peak at the cathode (\DriftTL) has a larger tail to the right in data that is not reproduced in simulation. This discrepancy is attributed to the detailed geometry and electrostatic properties of the plated source and cathode mesh which are not fully captured in the simulation.}
    \label{fig:driftTimes}
\end{figure}

\begin{figure}
    \centering    
 \includegraphics[width=\linewidth]{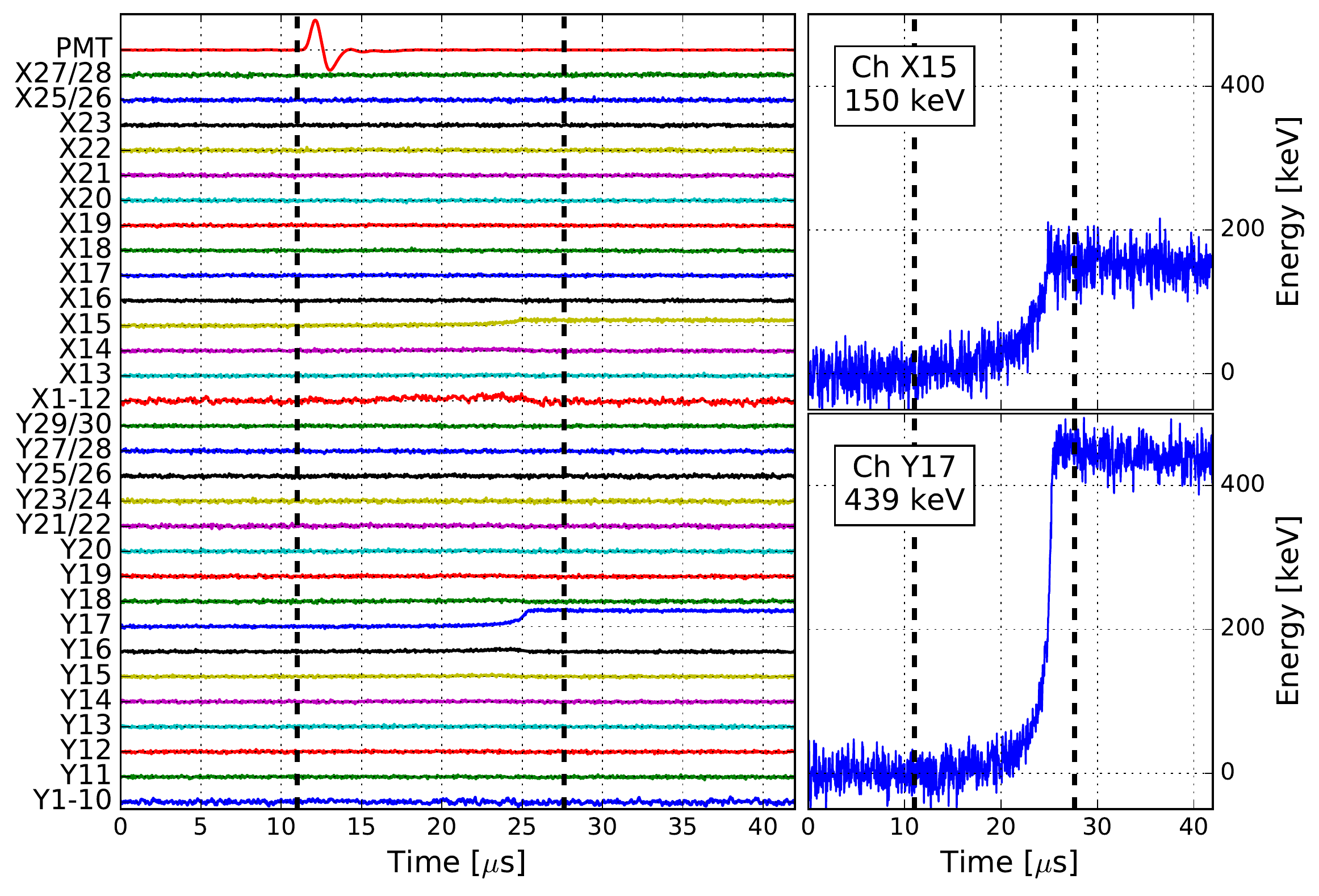}
 \caption{A sample event that passed all of the data selection cuts and fell in the Bi-207 peak at \BiPeakL. On the left the WFs on all charge channels and the PMT are shown, with the grouping of strips in each channel labeled on the vertical axis. Two single-strip charge channels received energy deposits above thresholds (X15 and Y17) resulting in a total energy deposit of 589~\si{keV}. The vertical lines at 11~\si{\us} represents the trigger from the PMT. The vertical line at 27.6~\si{\us} is the maximum expected drift time for an event that started at 11~\si{\us}.  On the right is a zoomed in view of the 2 charge channels above threshold.   Refer to Figure~\ref{Tilefig1} for locations of the channels.}
\label{WFfig}
\end{figure}

At a field of \Efield, the drift velocity of electrons in LXe is approximately \DriftV as measured in \cite{Miller:DriftV}. This gives a maximum drift time of \DriftTL at the drift length of \DriftL.  This is later confirmed by comparing the observed peak drift time in data to that in simulation, as shown in Figure~\ref{fig:driftTimes}. The digitized waveforms are 42~\si{\us} long (1050 samples) and include 11~\si{\us} (275 samples) before the PMT trigger. A sample event is shown in Figure~\ref{WFfig}, which also shows how strip channels are grouped for readout.

\section{Data Analysis}
\label{sec:analysis}

\begin{figure}
    \centering
    \includegraphics[trim={1.85cm 0 2.5cm 1cm},clip,width=1.0\linewidth]{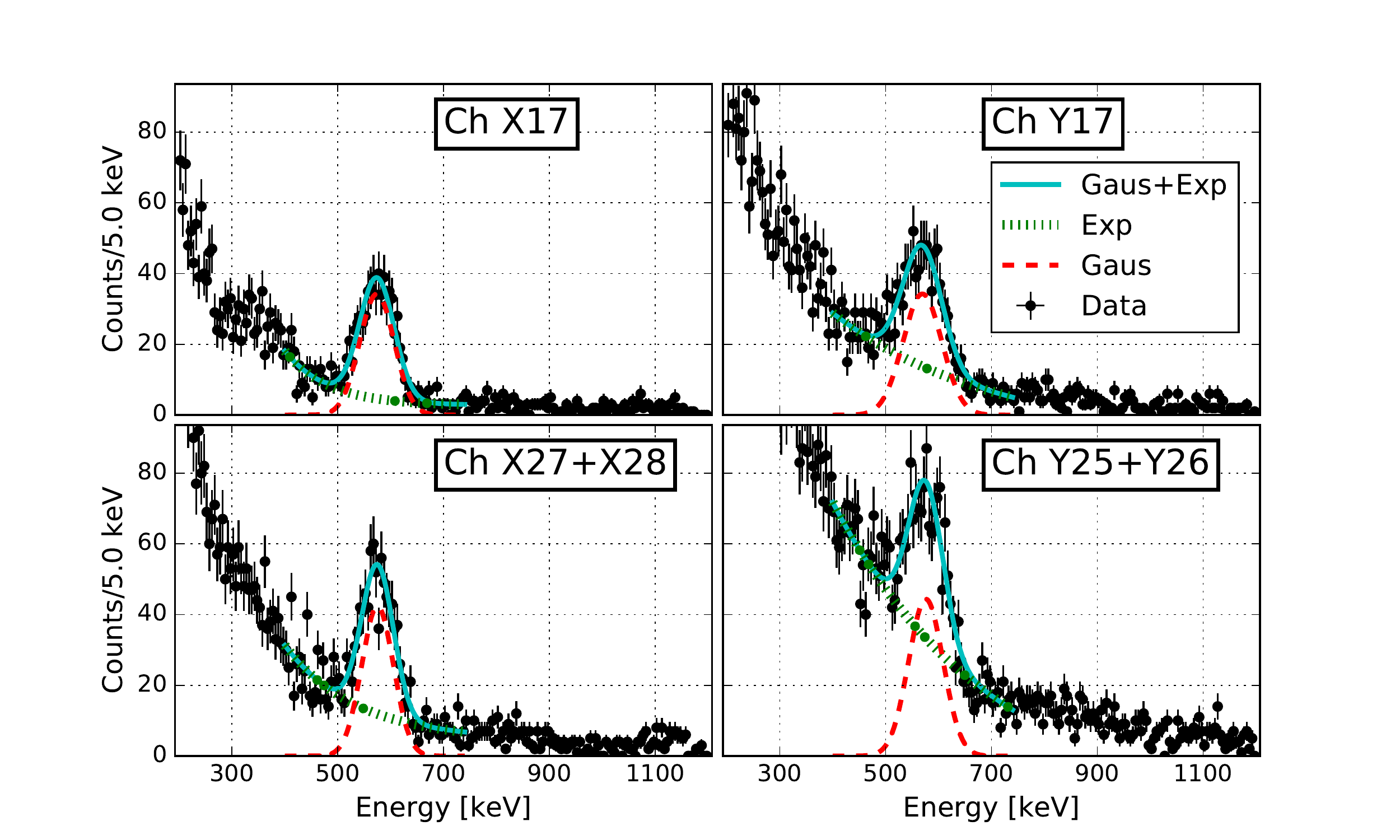}
    \caption{Example energy calibration fit to the \BiPeakL peak for single strip channels X17(top-left) and Y17(top-right) and double strip channels X27+X28 (lower-left) and Y25+Y26 (lower-right). A drift time cut and single-channel cut is included to select events from data. The full fit (solid blue) to the \BiPeakL for each channel is included as well as the individual Gaussian (dashed red) and the Exponential (hashed green) components which make up the full fit. }
    \label{fig:calibrationFits}
\end{figure}

\begin{figure}
    \centering
    \includegraphics[width=\linewidth]{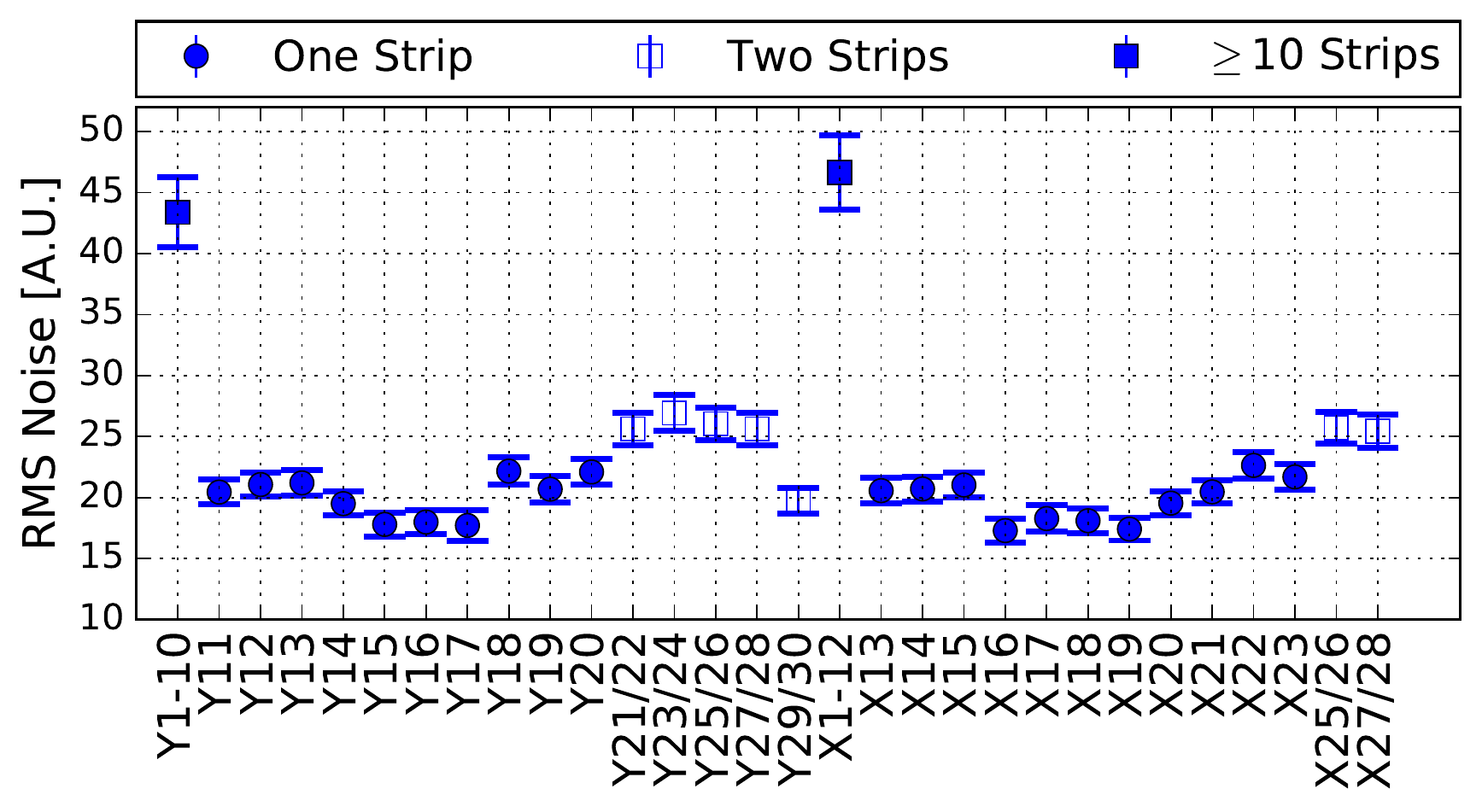}
    \caption{RMS noise, measured from 275-sample baseline, for each charge readout channel. Each data point is the mean of the RMS distribution; the error bars are the RMS width of the RMS distribution.  Channels with more than one strip have, on average, larger RMS noise because of their larger capacitance.  In general, the RMS noise grows with the square-root of the number of strips ganged together in the readout channel.}
    \label{fig:rmsVsCh}
\end{figure}

Once acquired, the digitized data is processed to extract parameters such as waveform energy, PMT amplitude, pulse rise times and delay between PMT trigger and charge collection. Each event contains 30 charge-channel waveforms.  Data analysis to extract a good quality energy measurement proceeds in two phases.  First the output of each channel is analyzed separately to look for individual signals. Depending on the results of this first pass, channels are then grouped together to determine the total energy deposited in the LXe. In the initial signal finding stage each charge-channel waveform is processed as follows:

\begin{enumerate}
\item The first 275 samples of each charge waveform, which are before the trigger, are used to compute the mean and RMS of the waveform baseline. The mean baseline value is then subtracted from the waveform.

\item The waveform is corrected for the exponential decay of the signal, using time constants measured for each charge-sensitive preamplifier. The decay times of the preamplifiers are long compared to the trace lengths, ranging from 200 to 500~\si{\us}, making this corrections small.

\item The last 275 samples of the waveform are averaged. Since the baseline has already been subtracted, this plateau is the uncalibrated charge energy, in ADC units. 

\item The energy is multiplied by a calibration constant determined from an energy spectrum fit, examples fits for a few channels can be seen in Figure~\ref{fig:calibrationFits}.

\item An energy threshold is applied by selecting channels with a measured energy that is more than five times the baseline RMS determined in Step~1. The average RMS noise of each channel is shown in Figure~\ref{fig:rmsVsCh}. The energy of all channels above this threshold are added to produce a total event energy. 
\end{enumerate}

The energy calibration for each charge channel is determined from fit to the highest intensity peak in the $^{207}$Bi spectrum, at \BiPeakL. For the individual channel calibration a single-channel cut is applied selecting only events in which only one channel is hit.  This ensures that there is no missing energy  on the channel being calibrated, at least up to deposits buried beneath the RMS noise of neighboring channels. In turn, this selection produces the most prominent peak in each channel's energy spectrum. The \BiPeakL peak is fit with a Gaussian function added with an exponentially-decaying background. An example fit is shown in Figure~\ref{fig:calibrationFits}.

After waveforms from each calibrated channel in an event are processed, some event-level information is calculated. The energy from all channels above the energy threshold are added to determine the total event energy. The waveforms above threshold are added to produce a sum waveform. The 95\% rise time is calculated for the sum waveform, and is used as a measure of the drift time. Cuts are applied to exclude events if they meet any of the following criteria:
\begin{enumerate}
\item Any waveform reaches the maximum or minimum of the ADC range. The range of all charge channels are $>7$~MeV, which is well outside the range of energies studied here.
\item The PMT waveform shape has poor agreement with a template shape. This cut eliminates pileup events where, {\it e.g.}, two decays of $^{207}$Bi occur close in time.
\item In any channel, the RMS noise computed with the 275 waveform samples used for either baseline averaging or energy measurement has a 5~sigma or greater excursion from the usual distribution of values. This cut eliminates noise bursts and events where the preamplifier or amplifier are at its maximum value.
\end{enumerate}
In total these data quality checks remove $\sim$10$\%$ of the total triggers included in the data set analyzed here.

The calibrated energy spectrum of the $^{207}$Bi source, compared to the simulation, is shown in Figure~\ref{fig:spec}. The production details of the simulated events is discussed in the next section.  The following cuts were applied to the spectra of both data and simulation to avoid pathological events either due to the small dimensions of the TPC or which are not properly modeled by the simulation:

\begin{figure}
   \centering
   \includegraphics[trim={0.8cm 0 2.3cm 0.5cm},clip,width=\linewidth]{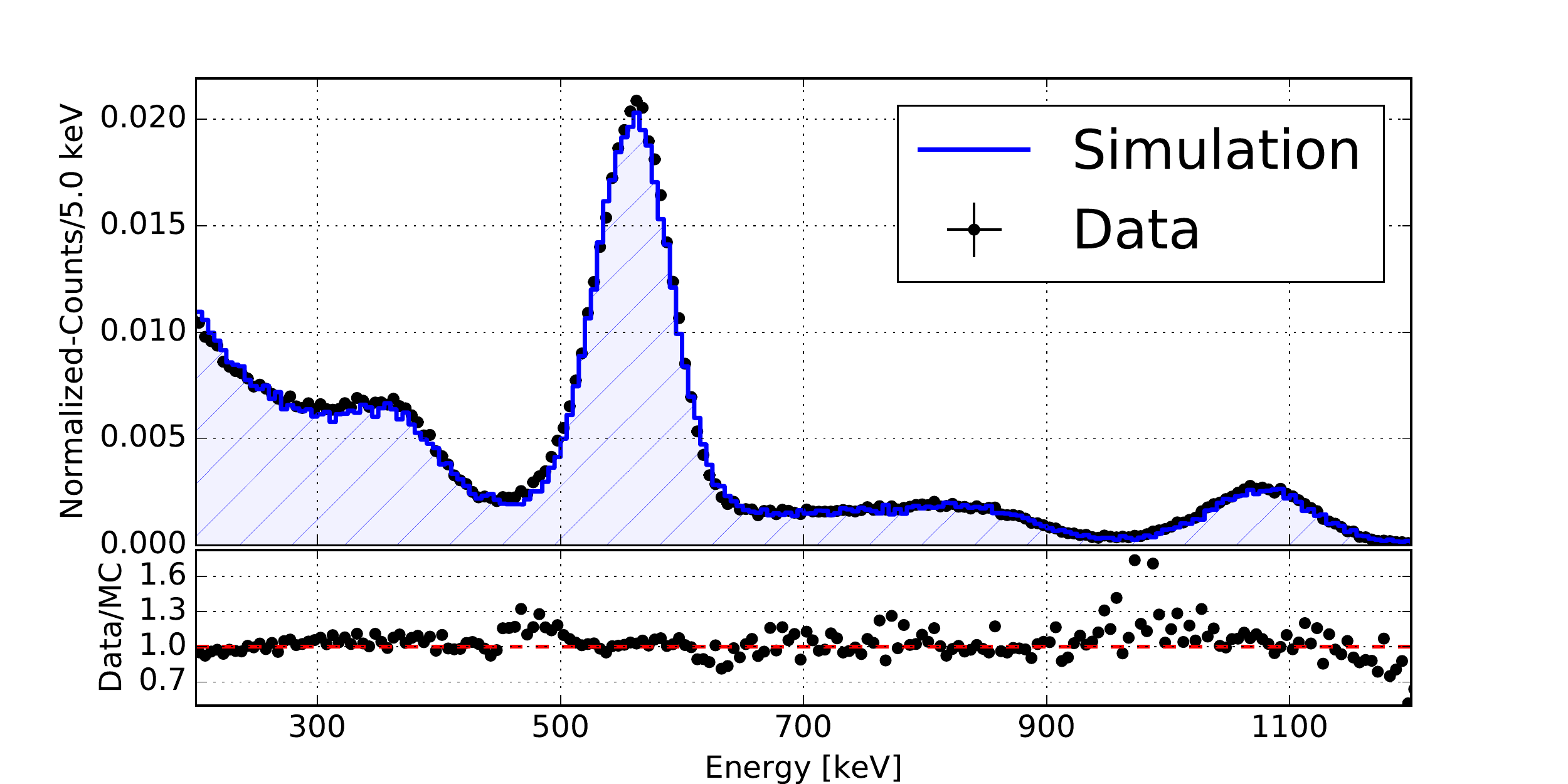}
   \caption{Energy spectra from data and simulation. Cuts are described in the text. The simulated spectrum was normalized to match the counts in the data spectrum.  The lower panel shows the ratio of the data and MC spectra with a horizontal line in red to show unity.}
   \label{fig:spec}
\end{figure}

\begin{figure}
   \centering
\begin{subfigure}{.49\textwidth}
  \centering
  \includegraphics[trim={1cm 0 2cm 0.0cm},clip,width=1.0\linewidth]{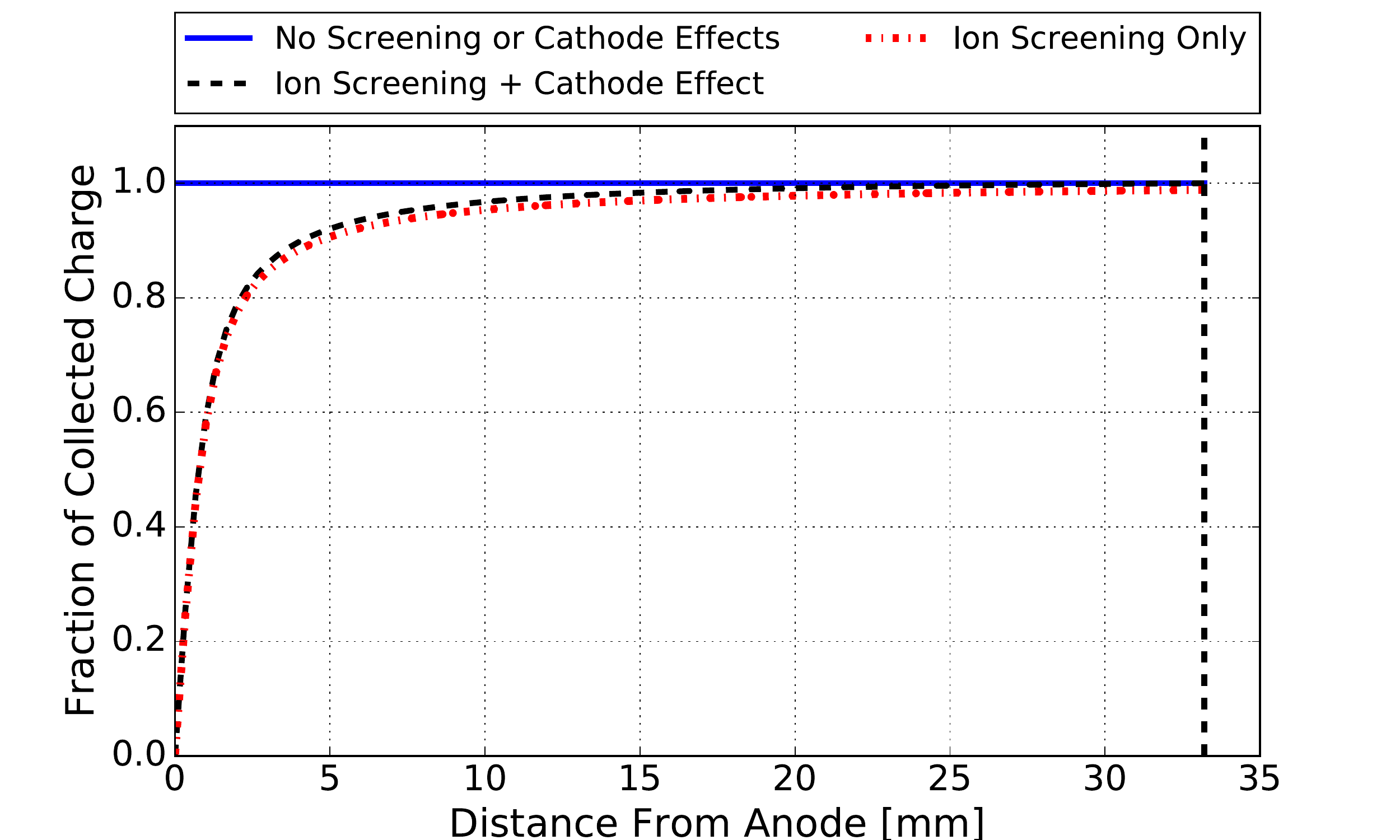}
  \caption{}
  \label{fig:electrostaticEffects_corr}
\end{subfigure}
\begin{subfigure}{.49\textwidth}
  \centering
  \includegraphics[trim={1cm 0 2cm 0.0cm},clip,width=1.0\linewidth]{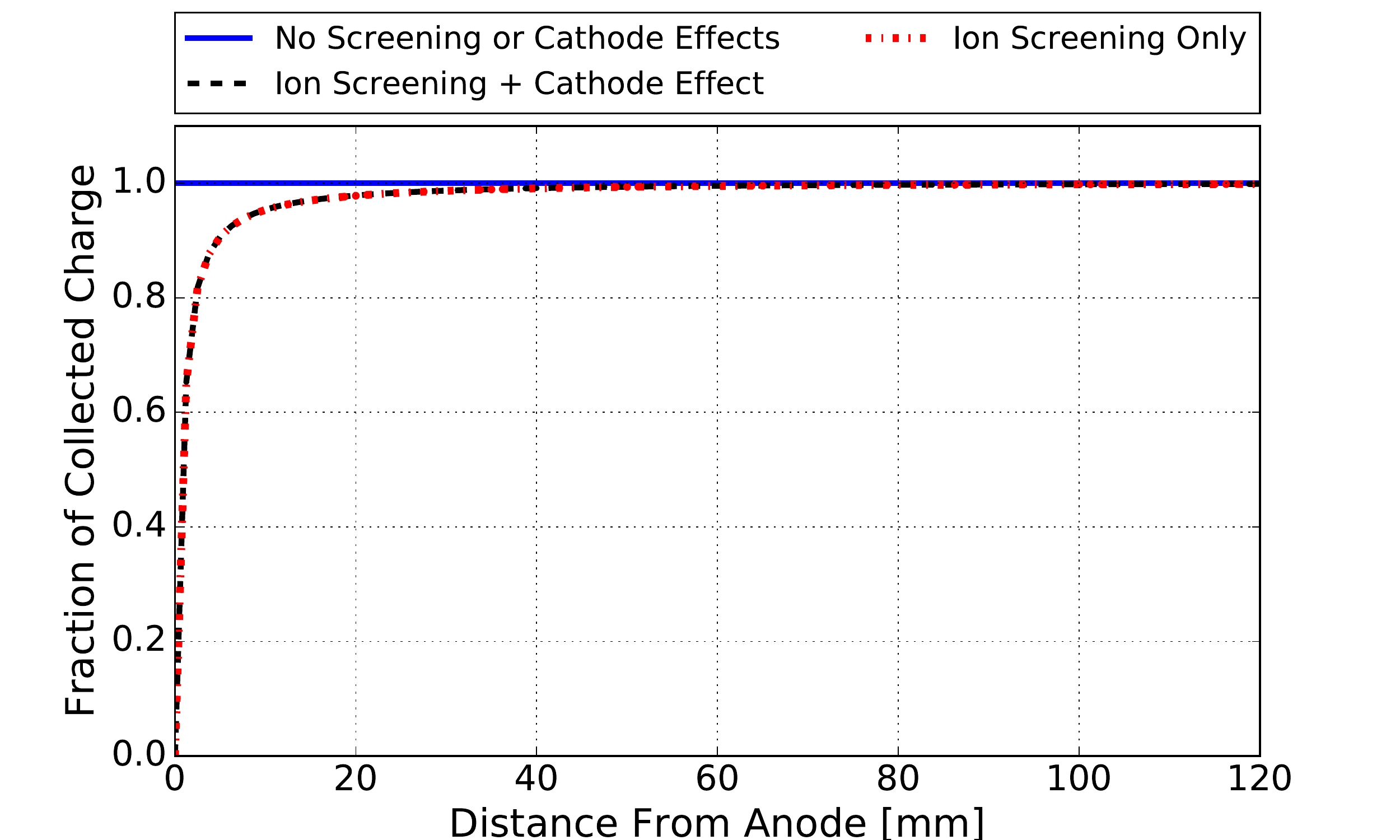}
  \caption{}
   \label{fig:nexoWP}
\end{subfigure}
\begin{subfigure}{\textwidth}
  \centering
  \includegraphics[trim={0cm 0 0cm 0cm},clip,width=1.0\linewidth]{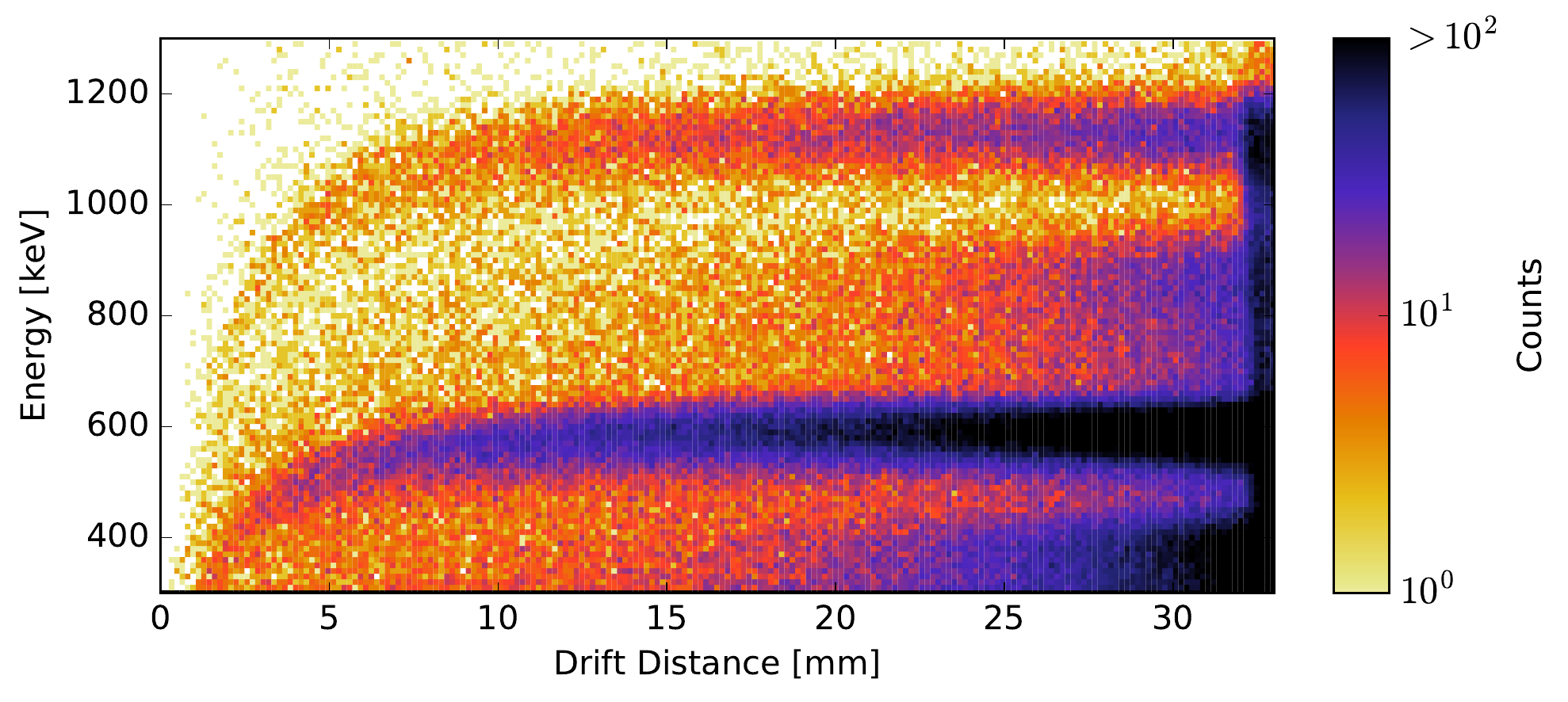}
  \caption{}
   \label{fig:data_electrostatic_effects}
\end{subfigure}
    \caption{Summary of the electrostatic effects (a) in full volume of this test setup (\DriftL drift length) and (b) the first 120~\si{mm} of a $\sim$1.3~\si{m} long detector such as nEXO.  The vertical axis shows the magnitude of the signal induced on one strip in response to an energy deposit directly below the strip. This is shown for the hypothetical case where the detector response does not vary with distance from the anode (solid blue) as well the case were only effects of ion screening (dashed black) and ion screening and cathode suppression (hashed red) are considered. Details of these effects are summarized in Appendix~\ref{sec:appendixA}.  Also shown (c) is the reconstructed energy vs drift distance observed in data where the predicted decrease in reconstructed energy close to the anode is observed. }
   \label{fig:electrostaticEffects}
\end{figure}

\begin{itemize}
\item The event is required to hit exactly one X channel and exactly one Y channel, to ensure that its location is well known in the TPC. This eliminates events happening in the xenon volumes above the ceramic boards, which are not described by the simulation.  This also removes events which have a light signal but no charge collected, which occurs when events interact with the LXe behind the cathode.  This should roughly occur $\sim$50$\%$ of the time since the source emits radiation in all directions. 

\item The X and Y channels are required to be single-strip readout channels. The single-strip channels have the best energy resolution since they have the smallest area and correspondingly smallest capacitance. As a result of the chosen wiring map, this requirement constrains the event to be near the center of the detector, far from the ceramic boards were the drift field is less uniform and energy can be lost to charge depositing on the ceramic instead of the tile (see Figure~\ref{Tilefig1}). This cut, along with the number of signals cut, remove $\sim$88$\%$ of the remaining events which passed the data quality cuts. 

\item The time delay between the PMT trigger and the arrival of the charge event is required to be >9.5~\si{\us}. This cut rejects events closer than 19~mm from the anode, where the signal height can vary by more than 1\% due to electrostatic effects described in Section~\ref{sec:mc2} and Appendix~\ref{sec:appendixA} (see Figure~\ref{fig:electrostaticEffects}).  This avoided introducing additional errors from imperfect modeling of these effects.    

\item The time delay between the PMT trigger and the arrival of the event is required to be $\le$16.1~\si{\us}. Since the maximum expected drift time is \DriftTL, this cut rejects energy deposits within 1~mm of the cathode, where the electric field is distorted, an effect attributable to the 3~mm spacing of the cathode mesh. This cut also eliminates pileup events (see Figure~\ref{fig:driftTimes}).   
\end{itemize}
The drift time cuts remove an additional $\sim$56$\%$ of the total events that passed the other cuts.  In total $\sim$5$\%$ of the total data included in the full data set is left for the final analysis.  

\section{Detector Simulation}
\label{sec:mc}

The simulation of the detector is split into two independent stages.  The first stage uses a GEANT4-based application \cite{G4Paper,Agostinelli2003250} in addition to the Noble Element Simulation Technique (NEST) model \cite{NEST_SIM} to parametrize the geometry and determine the number and location of ionization electrons and scintillation photons produced by particles interacting in the detector.  The version GEANT4.10.2.p02 is used for this stage of the simulation. The second stage of the simulation uses the output of the first stage to simulate the signals produced by drifting electrons collected on the strips and the electronics response of the detector.     

\subsection{GEANT4/NEST Simulation}

A detailed description of the detector geometry is input into a GEANT4 application, including the TPC Vessel, the PMT window, the internal components (tile, interface boards, cathode, the source, etc.), and the surrounding HFE cryostat.    

\begin{figure}
   \centering
  \includegraphics[width=\linewidth]{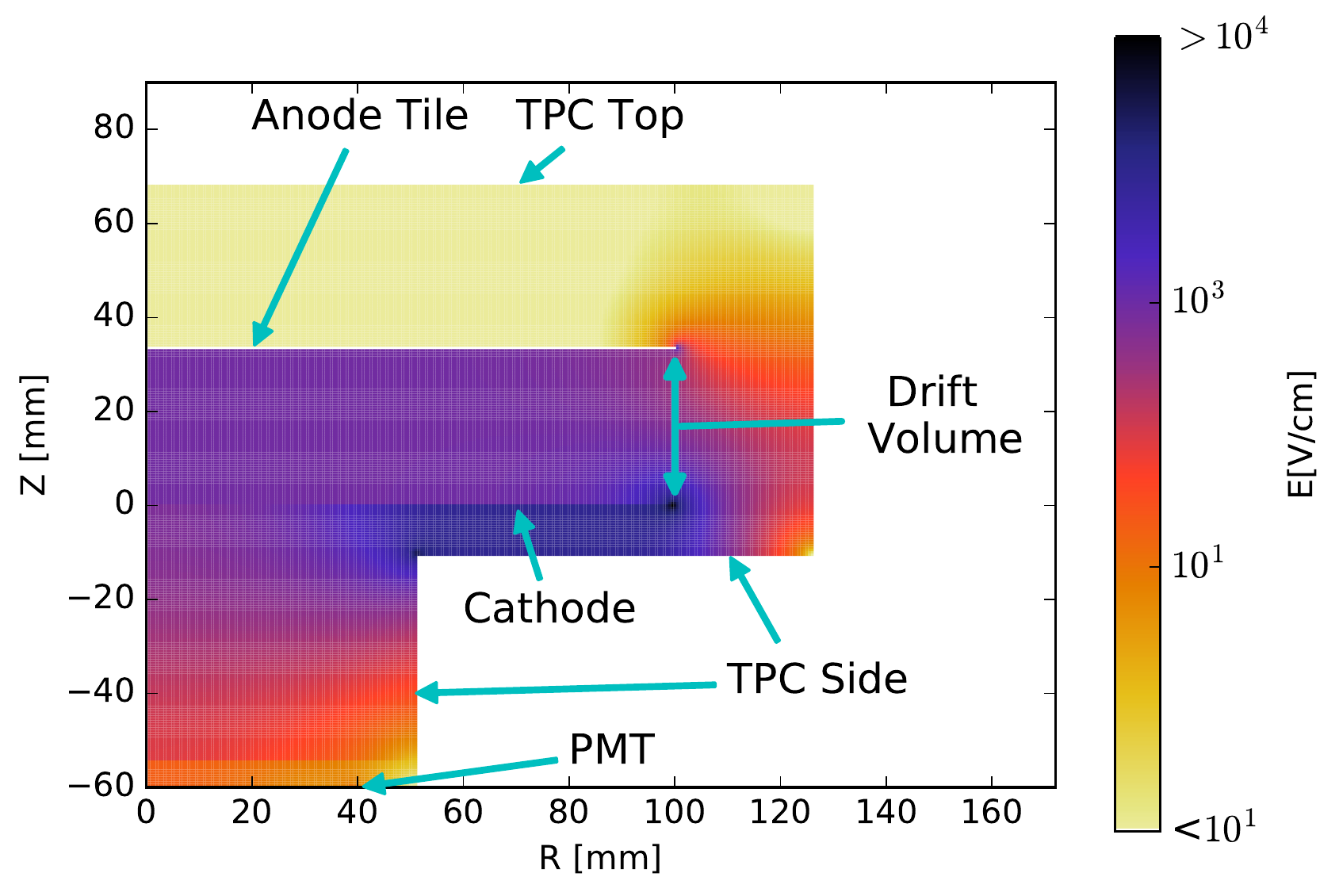}
   \caption{Azimuthally-symmetric map of the electric field in the liquid xenon volume. The cathode is at z=0 and the anode is at z=\DriftL. The cathode is biased at \CathBias while the anode and the TPC walls are at ground.}
   \label{fig:efield}
\end{figure}

Particles interacting with the LXe deposit energy by producing both scintillation light (178 nm) and electron-ion pairs (ionization).  Due to recombination of a fraction of electron-ion pairs the charge and light yield of individual events of a given energy and drift field display anti-correlated fluctuations that conserve the total deposited energy~\cite{Conti:2003av}.  The microphysics that determines the relative amount of energy going into each channel is not currently supported by the standard GEANT4 package.  To model this response NEST was used to accurately simulate both the ionization and scintillation response for different detector configurations.  The relative light and charge yield varies with the local electric field in the region of the energy deposit, which in turn determines the amount of recombination.  To account for this, a map of the electric field magnitude was produced from a cylindrically symmetric COMSOL~\cite{COMSOL} simulation of the liquid xenon volume of the detector.  This electric-field map is used by the simulation to provide the correct magnitude of the electric field to NEST at each point in space. The field map, which neglects the complicated geometry of the cathode mesh by modeling as a flat sheet, is shown in Figure~\ref{fig:efield}. The drift field in the region under the tile, assumed to be circular, varies from 890~V/cm to 990~V/cm.  This results in an average yield of $\sim$25~k electrons and $\sim$15~k photons for gamma's in the \BiPeakL peak.

Since the cuts described in Section~\ref{sec:analysis} constrain events to be in the center of the detector, only the $^{207}$Bi source located under the center of the anode was simulated to produce the energy spectrum for the single-strip channels shown in Figure~\ref{fig:spec}.  The complicated geometry of the $^{207}$Bi source is approximated as a circular disk of radius 2~mm in the plane of the cathode.  

\subsection{Signal Simulation}
\label{sec:mc2}
Ionized electrons produced in the first stage of simulation that fall within the drift region of the detector are diffused radially according to their drift time, using an electric-field-dependent transverse-diffusion coefficient determined from~\cite{EXODiff}.  Currently no diffusion in the longitudinal direction is included in simulation.  The diffused electrons are then binned into voxels with sides of length 530~\si{\um} in the x and y directions and 80~\si{\um} along the drift direction, z.  The x and y dimensions of the voxels were chosen to minimize processing time while preserving signal quality; the z dimension is equivalent to one sample of the 25~MS/s digitizer for a drift velocity of \DriftV. Each voxel is  tracked as it is drifted from the interaction location to the charge collection tile assuming a uniform drift velocity along the drift axis. The approximation of perfectly parallel field lines is valid in the bulk of the LXe but not near the cathode, where the non-uniform geometry of the mesh as well as the source wire shown in Figure~\ref{fig:dewar} cause field distortions.  This results in some divergence in the field lines and subsequently the path of electrons, broadening the drift time of events in data near the cathode.  From Figure~\ref{fig:driftTimes} this smearing occurs as expected near the cathode where the data sees a much broader peak of maximum drift times.  This disagreement between data and simulation motivates the fiducial cut to remove events near the cathode. Diverging field lines are also expected near the edges of the drift region but events in those regions are rejected from the current analysis which only looks at events which hit central strips.  For every time step of 40~\si{ns} (0.08~\si{mm}), the charge induced on each readout strip is determined using an analytical calculation of the charge induced on a square pad in an infinite plane electrode arrangement.  The charge induced on a single strip is calculated as the sum of the induced charge on each pad comprising the strip.  Corrections are applied to account for electrostatic effects inside the TPC that affect the development of the charge signal, such as cathode suppression and ion screening.  A summary of these effects is shown in Figure~\ref{fig:electrostaticEffects} and the details of the calculation are described in Appendix~\ref{sec:appendixA}.

Waveforms for each charge readout channel are then produced using a charge propagation simulation to track electrons from their initial deposition location to their final collection point.  The waveforms are sampled at 25~MS/s and contain 1050 samples each, to match the data measured from the LXe setup.  In order to make simulated waveforms more realistic, noise waveforms are recorded with solicited triggers throughout data taking.  A sample of these solicited waveforms is superimposed on those generated in simulation from drifting charge resulting in waveforms that more accurately resemble what is recorded during data collection.  Simulated charge waveforms are then processed using the same reconstruction algorithm as data to find signals and determine event energies.  The effect of the electronics used here (Figure~\ref{fig:preamp}) on the waveform shape is negligible and is not included in the simulation. 

Although the first stage of the simulation produces both ionization and scintillation signals, the latter is not propagated to the signal generation stage of the simulation since the light response of the detector is currently used solely as a trigger and not for measuring event energy.  This is the result of the single PMT being heavily shadowed by the layout of the charge tile and the amount of scintillation light it detects limits its use for an energy measurement. Work is currently in progress to use an array of high-QE, cryogenic SiPM to detect the LXe scintillation light and overcome this limitation.

\begin{figure}
    \centering
    \includegraphics[trim={1cm 0 2cm 0.0cm},clip,width=\linewidth]{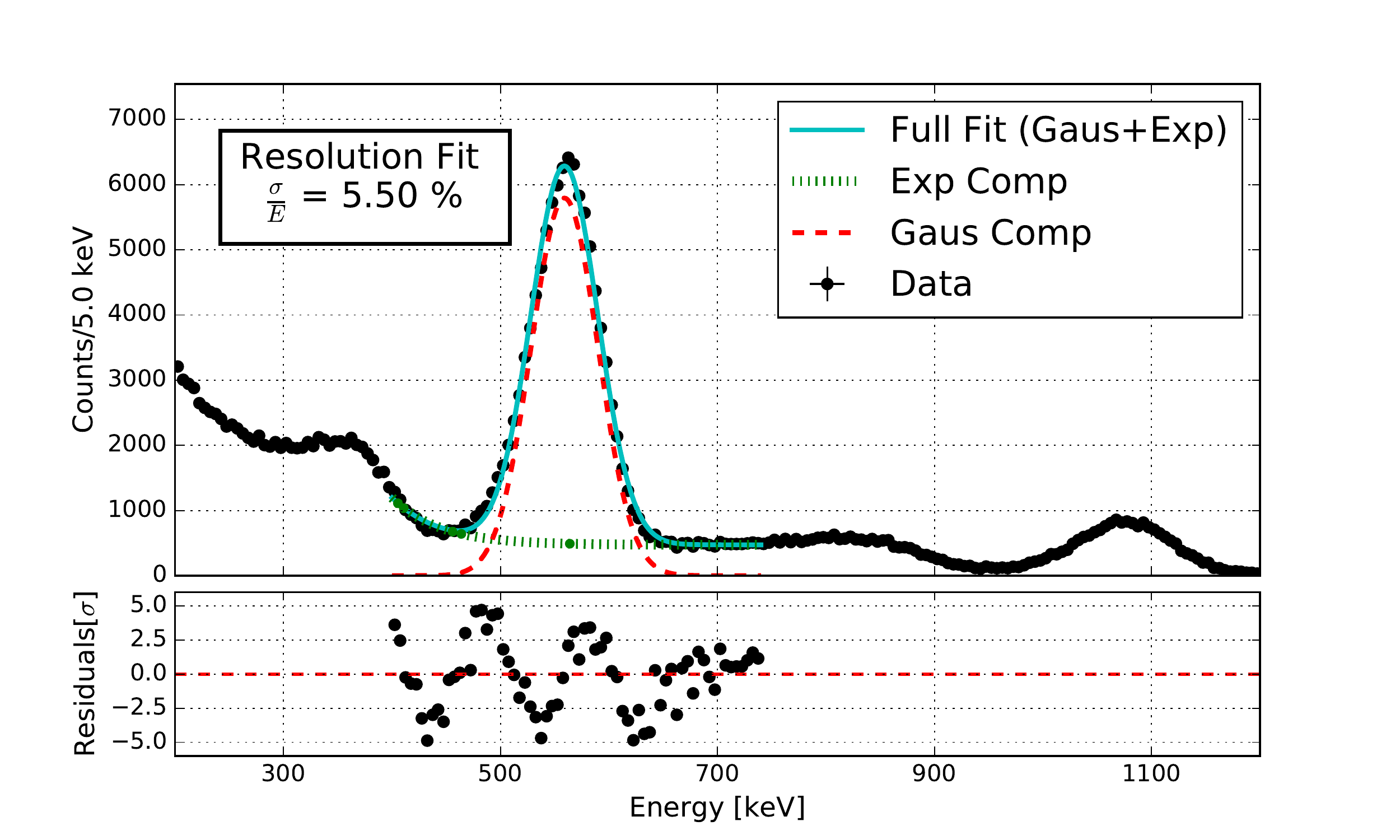}
    \caption{Fit to the \BiPeakL peak (solid blue) seen in data using the cuts described in the text.  The fitting function is composed of a Gaussian (dashed red) summed with an exponential background (hashed green) which are also shown. The structure in the residuals is due to the simplifying assumptions of the fit model and doesn't affect the measured energy resolution.}
    \label{fig:ResolutionFits}
\end{figure}

Current electronic simulations do not take into account any cross talk which may be present between crossed charge channels. Dedicated studies of this cross talk showed evidence of small levels of signal contamination.  The observed signals resulting from these cross talk studies were shown to mimic induction only like signals.   For the current analysis which only looks at collection signals the impact of cross talk is minimal, but the treatment of induction signals, particularly for a grid-less design, will be important for the future studies.

\section{Results}
\label{sec:results}

A comparison between the energy spectra from simulation and data is shown in Figure~\ref{fig:spec}. The spectra for both are normalized to have equal area in the energy range 200~\si{\keV} to 1200~\si{keV}.  At energies above 200~keV, the simulation is in good agreement with data. Below 200~keV, the data spectrum has fewer counts than predicted by simulations because of PMT trigger threshold effects. In addition Figure~\ref{fig:electrostaticEffects_corr} and Figure~\ref{fig:data_electrostatic_effects} show the predicted and observed electrostatic effects in the current detector as a function of drift distance respectively.  The trend of decreasing reconstructed energy near the anode plane is observed in both supporting the model used in simulation.
Fits to the \BiPeakL peak from data and simulation are shown in Figure~\ref{fig:ResolutionFits}.
The noise-subtracted ionization-only energy resolution of \DataRes at \BiPeakL is consistent with the intrinsic resolution of liquid xenon measured by other investigators~\cite{Conti:2003av,AprileDoke:2010Diff}. 

\section{Conclusions}
\label{sec:conclusions}

We report on the performance of a novel, segmented, grid-less ionization charge collection detector developed for the nEXO 5-tonne liquid xenon TPC for neutrinoless double beta decay. 
The charge-only energy resolution measured in LXe is in line with the intrinsic value measured for LXe by numerous investigators.  The data from the prototype "tile" shed light on non-conventional electrostatic effects arising from the absence of a shielding Frisch grid in front of the charge collection electrode. A study of these effects for a nEXO sized TPC ($\approx$130 cm drift length) are shown in Figure~\ref{fig:nexoWP} for a relative comparison to the same effects in the currently studied detector shown in Figure~\ref{fig:electrostaticEffects_corr}. This includes the position dependence of the reconstructed charge, including the ion screening and cathode effects described in Appendix~\ref{sec:appendixA}.

Work is in progress to refine the design of the charge tiles. The strip pitch is being optimized for use in nEXO. Future prototype tiles will run with integrated readout circuits placed in LXe. Finally, improvements in the light collection efficiency are being implemented to allow energy measurements taking advantage of the anti-correlation between the ionization and scintillation signals.

\acknowledgments{%
This work has been supported by the Offices of Nuclear and High Energy Physics within DOE's Office of  Science, and NSF in the United States, by NERSC, CFI, FRQNT, NRC in Canada, by SNF in Switzerland, by IBS  in Korea, by RFBR in Russia, and by CAS and ISTCP in  China. This work was supported in part by Laboratory Directed Research and Development (LDRD) programs at Brookhaven National Laboratory (BNL),  Lawrence Livermore National Laboratory (LLNL), Oak Ridge National Laboratory (ORNL) and Pacific Northwest National Laboratory (PNNL).
}

\appendix
\section{Induced Charge Calculation}
\label{sec:appendixA}

The induced charge on a single square pad comes from considering a point charge above a rectangular plane of grounded electrodes.  The induced charge per unit area on a conducting plane by a point charge of charge $Q_{0}$ located a distance h above the pad at x=y=0 is given by method of images as

\begin{equation}
\sigma  = \frac{-Q_{0}h}{2\pi(r^{2} + h^{2})^{3/2}}
\end{equation}
To find the charge in a rectangle which extends from $x_1$ to $x_2$ along the $x$-axis and $y_1$ and $y_2$ along the $y$-axis we calculate
\begin{equation}
Q(h)=-\frac{Q_0 h}{2\pi}\int_{x_1}^{x_2}\int_{y_1}^{y_2}\frac{dx\,dy}{(x^2+y^2+h^2)^{3/2}}
\end{equation}
which can be evaluated as
\begin{equation}
Q(x, y, h) = -\frac{Q_{0}}{2\pi} \left[ f(x_{2}, y_{2}, h) - f(x_{1}, y_{2}, h) - f(x_{2}, y_{1}, h) + f(x_{1}, y_{1}, h)\right]
\label{ChargeIndEqn1}
\end{equation}
\noindent where Q is the charge induced on the pad, $Q_0$ is the magnitude of the drifting change, the set of $x_{i}$ and $y_{i}$ are the distances from the drifting charge to the four corners of the pad, and $h$ is the height of the charge above the anode. The function $f(x,y,h)$ is defined below: 
\begin{equation}
f(x,y,h) = \arctan \left[ \frac{xy}{h\sqrt{x^2 + y^2 + h^2}} \right].
\label{ChargeIndEqn2}
\end{equation}
\noindent The value of $h$ decreases with time according to the drift velocity, \DriftV at \Efield. 

Two  corrections are applied to account for electrostatic effects from the positive Xe ions produced in the ionization process as well as the charge induced on the cathode.

The correction for the positive ion is an added term,  Equation~\ref{ChargeIndEqn1}, with the opposite sign to account for the positive charge and a constant value of $h$ since the ion is approximately stationary during the duration of the event.  The  correction for the cathode is a linear scaling to the induced charge based on the distance from the cathode. 

\begin{equation}
Q_{Full}(x, y, h) = \left(\frac{D-h}{D}\right)Q(x, y, h) - \left(\frac{D-h_0}{D}\right)Q(x, y, h_0)
\label{ChargeIndEqn3}
\end{equation}

\noindent where $D$ is the distance between the cathode and the anode, $h_0$ is the initial height of the event, and $h$ is a function of time. 

A third correction for a finite electron lifetime was considered but was ignored since no effects of purity degradation were observed. A plot showing the relative charge induced on the cathode for a charge at different z-positions in the detector is shown in Figure~\ref{fig:electrostaticEffects}, for the current test setup (left panel) as well as a $\sim$1~\si{m} detector such as nEXO (right panel). 

The full expression for charge induced on a strip is calculated by summing Equation~\ref{ChargeIndEqn3} over all 30 pads in a strip for each voxel of charge considered in the simulation.

\bibliographystyle{JHEP} 

\vspace{1cm}

\bibliography{references}

\end{document}

%% file: authors_full.tex
\newcommand{\Stanford}{1}
\newcommand{\IHEP}{2}
\newcommand{\SLAC}{3}
\newcommand{\UMass}{4}
\newcommand{\ORNL}{5}
\newcommand{\Yale}{6}
\newcommand{\Erlangen}{7}
\newcommand{\IME}{8}
\newcommand{\Indiana}{9}
\newcommand{\PNL}{10}
\newcommand{\Carleton}{11}
\newcommand{\Duke}{12}
\newcommand{\Illinois}{13}
\newcommand{\ITEP}{14}
\newcommand{\Sherbrooke}{15}
\newcommand{\LLNL}{16}
\newcommand{\RPI}{17}
\newcommand{\McGill}{18}
\newcommand{\Triumph}{19}
\newcommand{\Colorado}{20}
\newcommand{\BNL}{21}
\newcommand{\Laurentian}{22}
\newcommand{\SD}{23}
\newcommand{\Bama}{24}
\newcommand{\Drexel}{25}
\newcommand{\Stony}{26}
\newcommand{\IBS}{27}
\newcommand{\LHEP}{28}

\author[\Stanford]{M.~Jewell}
\author[\Stanford,a]{A.~Schubert%
\note[a]{now at OneBridge Solutions, Boise, ID}}
\author[\IHEP]{W.R.~Cen}
\author[\SLAC,\Stanford,\UMass]{J.~Dalmasson}
\author[\Stanford]{R.~DeVoe}
\author[\ORNL]{L.~Fabris}
\author[\Stanford]{G.~Gratta}
\author[\Yale,\Erlangen]{A.~Jamil}
\author[\Stanford]{G.~Li}
\author[\SLAC]{A.~Odian}
% ...

\collaboration{nEXO collaboration}